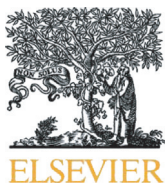
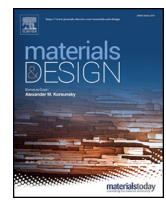

# A new approach to evaluate the elastic modulus of metallic foils

C.O.W. Trost [a,*], S. Wurster [a], C. Freitag [a], A. Steinberger [b], M.J. Cordill [a,c]

[a] Erich Schmid Institute of Materials Science, Austrian Academy of Sciences, Jahnstrasse 12, Leoben 8700, Austria
[b] Austria Technologie & Systemtechnik (AT&S) Aktiengesellschaft, Fabriksgasse 13, Leoben, Austria
[c] Department of Materials Science, Montanuniversität Leoben, Jahnstrasse 12, Leoben 8700, Austria

## HIGHLIGHTS

- The precise determination of the elastic modulus of metallic foils is non-trivial.
- Multiple elastic loading – unloading cycles led to reproducible and accurate determination compared to monotonic testing.
- Results were interpreted using Electron Backscatter Diffraction and X-Ray Diffraction methods.
- Foil cross sections were prepared and analysed.

## GRAPHICAL ABSTRACT

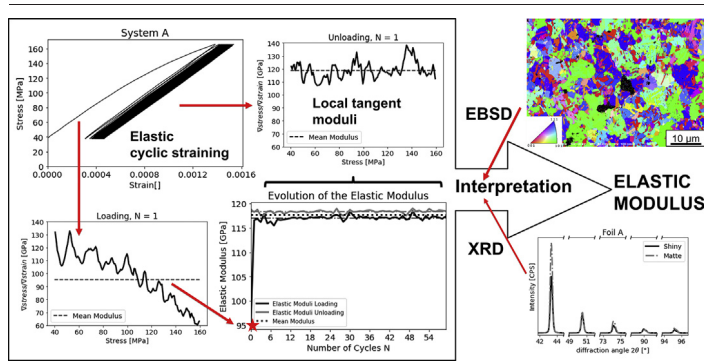

### ABSTRACT

The accurate determination of the elastic properties is non-trivial for metallic foils. The measured elastic modulus is often described in literature as significantly smaller than the respective modulus of the bulk counterparts. This paper describes a straight forward way to minimize the influence of the measurement on the calculated values using a standard tensile testing device combined with a laser speckle extensometer and advanced data evaluation. The elastic modulus obtained with a monotonic tensile procedure is compared to values obtained from multiple loading–unloading cycles in the elastic regime. The latter were found to lead to more reproducible and reasonable values with significantly smaller standard deviations. To enable interpretation of the modulus values electron backscatter diffraction and X-ray diffraction were used to emphasize the effects of texture and microstructure. Two different electrodeposited copper foils were characterized since copper is widely used in many industrial applications.

© 2020 The Authors. Published by Elsevier Ltd. This is an open access article under the CC BY-NC-ND license (http://creativecommons.org/licenses/by-nc-nd/4.0/).

## 1. Introduction

Accurately measuring the mechanical properties of metallic foils and ribbons is challenging. Due to their thicknesses and the tendency to deform while handling, foils need to be treated with care. There are several approaches to measure the elastic properties of foils, ribbons and films such as tensile testing [1–23], resonance method [3], dynamic mechanical analysis [4], nanoindentation [4,5,18,24], bulge testing [25–27] or texture based theoretical calculations [5,24,28]. It has been reported that the elastic modulus of metallic foils can be dependent on foil thickness, manufacturing method or microstructural factors [2,29,30]. One research group [3] showed that elastic modulus measured by straining varied compared to measurements made with the resonance method. The resonance method leads to elastic modulus values up to two times larger compared to values measured with the tensile test procedure of IPC TM-650 2.4.18.3 [3,31]. The differences between methods leads one to believe that a better testing procedure for foils is needed. Since tensile testing is common in most materials laboratories, emphasis will be on this technique.

* Corresponding author.
   E-mail address: claus.trost@oeaw.ac.at (C.O.W. Trost).

When tensile testing foils many factors that are straight forward when testing bulk samples can hinder the accurate measurement of mechanical properties. These factors include gripping problems, proper sample alignment between the grips, and sample production, all of which can be a challenge to overcome in the case of foils and seriously affect the outcome of the experiments. According to the ASTM standard (ASTM E345-16 [32]), two sample geometries are recommended for metallic foils: strips or dog bone. Strips can be produced by shearing or slitting, dog bones by milling-type cutting. Production by shearing or slitting can have various impacts on the measured data because production might leave a lip on the samples. The simple strip geometry can also lead to stress concentrations at the gripping area which result in fracture at the grips and an invalid experiment.

It is evident that the accurate determination of strain is crucial for determination of elastic properties. It has been shown that for foils > 100 μm the strain can still be measured by contact extensometers [33]. For thinner samples measuring the strain can be challenging since standard clip-gauges cannot be used and the crosshead movement overestimates the strain in the elastic regime and is not accurate enough [2,11,19]. This overestimation has various impacts not only lowering the calculated modulus, but also changing the elastic to plastic yield transition, which leads to an imprecise definition of the yield as described by [2]. Therefore, non-contact techniques like laser speckle extensometer (LSE) and digital image correlation (DIC) are now widely used to accurately measure the strain of foils and ribbons [2,13,19,29]. The initial straining data of foils and films often show artefacts and lack a clear linear elastic part. Hwangbo and Song [9] described that the straight, elastic (rectilinear) segment of the stress–strain curve of thin films is often unclear or only limited to a very small segment, which makes evaluation of an accurate elastic modulus difficult, ultimately leading to inexact determination of the 0.2% offset yield strength. Read [11] also described the initial response to loading of 2.6 μm electrodeposited (ED) copper specimens as typically nonlinear. To counter the described problems, various experiments for determination of the elastic modulus of freestanding films and foils have been proposed ranging from unloading one or multiple times during a complete tensile test [2,6–8,12,13,19–23] or using special cyclic loading tests [9–11] for evaluation. Cyclic loading was applied due to the fact that parts of the stress–strain curve were found to show ideal elastic behaviour and the elastic modulus for 12 μm electrodeposited copper foil measured by cyclic loading were significantly higher than the values measured using a monotonic tensile tests [9].

In this paper, a nanoindentation inspired [34] cyclic loading-unloading procedure to accurately measure and evaluate mechanical properties is proposed. The elastic loading and unloading portions of the test will be used to determine the elastic modulus. In contrast to nanoindentation, the global rather than the local elastic modulus can be measured. Furthermore, a way to find the optimal stress range for evaluation will be also proposed. The experiments were performed on two ED copper foils with thicknesses of ~40 μm having different microstructures and chemistries. Copper foils were studied because they have important industrial applications, thus making it crucial to determine mechanical properties accurately, in order to predict lifetimes and enable accurate simulations. To complement the analysis of the tensile tests, electron backscatter diffraction (EBSD), backscatter electron imaging (BSE), energy-dispersive X-ray spectroscopy (EDS), and X-ray diffraction (XRD) were used to fully characterize the materials. Ultimately, the tensile test data were processed by Python3 scripts to enable insights into the metadata of the performed tensile tests.

## 2. Experimental

Two industrially manufactured ED Cu foils were investigated (Foil A and Foil B). Due to the production process, the foils have two surfaces that differ in terms of appearance and roughness. This is a result of the growth of the foils on a cathode out of a chemical bath. ED foil surfaces generally can be distinguished as a shiny side facing the cathode and a matte side which faces the bath during deposition as also described by [5,9]. The two foils were tested weeks after deposition to ensure that the microstructure was completely self-annealed.

A scanning electron microscope (SEM, LEO 1525 Zeiss Inc., Oberkochen, Germany) was used to obtain SEM images and EBSD analysis was performed with a Bruker e⁻Flash$^{FS}$ detector of the microstructures of the foil surfaces and cross-sections. EBSD scans of the foils were made on samples that were electro-polished using a Struers TegraPol-5 from the matte side resulting in an average thickness of 28 μm. The used parameters were 35 V, a flowrate of 13 for 10 s and the used electrolyte (Struers - Electrolyte D2) was kept cool in a freezer before use. Producing cross sections was found to be challenging, and different mechanical, electro-polishing, and ion polishing methods were explored to find the optimal way of production as similarly described in [35,36]. Ultimately the cross sections for EBSD analysis were prepared with an ion slicer (Hitatchi Ion Milling System E-3500) using Ar ions with 6 kV acceleration voltage and 4 kV discharge voltage for 5 h. The EBSD scans were performed with a step size of 100 nm. For grain size analysis, the grains were defined by a maximum tolerance angle of 15° and at least 5 pixels per grain. EDS was performed with the same SEM using a Bruker XFlash 6|60 detector to qualitatively measure the foil chemistry. An acceleration voltage of 18 kV was taken and two copper peaks were identified properly. The traces of the elements sulphur and iron were also observed. The EBSD and EDS data analysis were performed with the Bruker Esprit 2.2 software. For X-ray diffraction, a Rigaku SmartLab 5-Axis X-ray diffractometer, with a CuKα wavelength in parallel beam mode, was used. Both sides of the foils were examined to determine the differences in texture resulting from the deposition process.

The samples for tensile testing were made according to ASTM E345–16 [32] using specimen shape A (dog bone shape) (Fig. 1a). Two production techniques were investigated, leading to two sets of samples. One set was produced by laser cutting using an industrial UV/C0$_2$ laser drilling machine. The second set of samples were cut with a steel carbide blade using the Cricut Explore Air 2 which is a state if the art method to produce samples for various experiments from foil/sheet materials (metals and polymers) [37–40]. The laser-cut samples (Fig. 1c) show a very fine and repetitive pattern at the sample edges and remain completely flat. The blade cut samples show a large deformed lip (Fig. 1b) indicating that the blade-cut samples are not as flat as perceived, adding unavoidable further measurement artefacts or even cutting defects leading to premature failure during testing. Furthermore, the samples easily deformed when unmounting from the adhesive Cricut cutting mat resulting in a loss of samples even before testing, making it an impractical method to produce multiple identical samples from needed for statistical evaluation from metallic foils. Ultimately, only the laser-cut samples were used for evaluation in this study. The actual thicknesses of the foils were measured using a foil thickness gauge by taking the mean of at least six measurements outside the gauge area before an experiment was performed.

The laser-cut foils were strained in uniaxial tension using a ZWICK/ROELL Z100 equipped with 200 N Load Cell type ZWICK/ROELL Xforce P with a sensitivity of 2 mV/V combined with flat grips ZWICK/ROELL Type 8033. To ensure proper gripping, the grips were sand blasted and pre-tests were performed to ensure no slipping occurred. A LSE detector ZWICK/ROELL laserXtens HP TZ was used to measure strain of the samples. A gauge length of 50 mm was used. The tensile device and the LSE detector were controlled with the testXpert III–V1.2 software. To enable proper alignment of the LSE detector a preload of 20 N was used. The tests were performed using force control always going up to 87.5 N and down to 20 N. A crosshead speed of 0.6 mm/min was used for all experiments. Only samples without artefacts in the cyclic regime were considered. Six samples per foil type (A and B) were evaluated with 60 load-unload cycles in the elastic regime per test. For each foil type, the elastic modulus was determined by calculating the average for different combinations of unloading and loading segments of the engineering stress–strain curve, with the first loading segments being markedly different, as will be shown later. The evaluation was performed in different stress regimes, and the plateaus of the modulus values between 80 and 140 MPa were most stable as will be demonstrated later. This is valid for both the loading and unloading cycles.

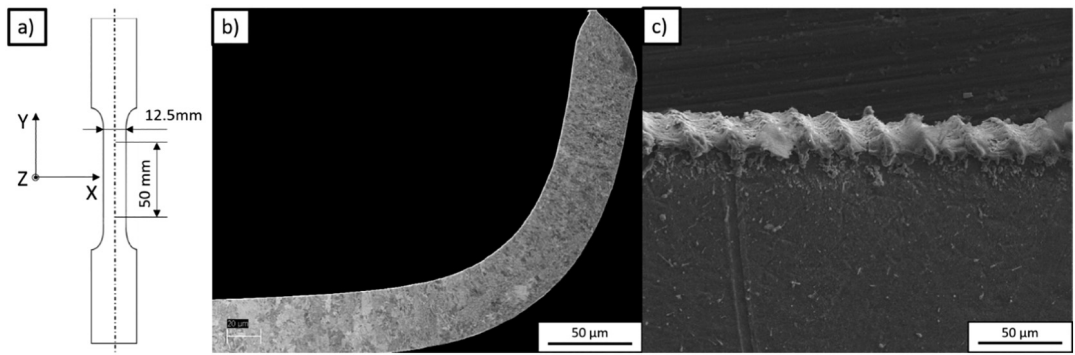

**Fig. 1.** a) Dog bone schematic, for detail see [32]. b) BSE cross sectional image of a sample cut with a steel carbide blade showing the lip at the cutting edge. c) Repetitive pattern obtained by laser cutting at the edge of the Cu foils.

## 3. Results

### 3.1. Foil chemistry and microstructure

Fig. 2 shows that there is a distinct difference between the surfaces of the two systems. For both sides, the shiny (drum, cathode) side and the matte (bath, anode) side, the morphology is a direct result of the chosen deposition parameters. Foil A shows fine structures on the matte side while Foil B shows a coarser structured surface.

EDS scans showed that Foil A was nearly impurity free (within the limit of detectability of EDS) and Foil B was found to contain sulphur impurities. Sulphur can be considered a residual of the electrodeposition process [41,42]. Shiny and matte sides of both foils showed minor differences in terms of chemical composition. Furthermore, traces of iron could be found on the shiny side. This can be a direct result of deposition process.

The EBSD scans in Fig. 3 of the electro-polished matte side of the foils show differences in terms of grain size, grain distribution, and twinning behaviour. When analysing Sigma 3 ⟨111⟩ grain boundaries one can see that Foil A shows more pronounced twinning with 0.99 μm/μm$^2$ for a processed area of 2500 μm$^2$, Foil B has 1.28 μm/μm$^2$ for the same area. Foil A has larger grains (Fig. 3a) compared to Foil B (Fig. 3b). Fig. 3a and b illustrate differences in overall orientation of the grains and that there is a slight 111 texture in Foil A, but no strong texture observed for Foil B (Fig. 3c and d). Further microstructural differences can also be seen in the grain size distribution curves (Fig. 4). The grain size distributions were plotted in diagrams inspired by sieve curves commonly known in ceramics and powder metallurgy fields. The used plots differ from both classical sieve analysis inspired curves already used in the EBSD field [43], because a logarithmic scale is not used and from grain size distribution curves used commonly in EBSD by accumulating the grain size distribution. The curves were already used to describe EBSD scans of copper [22]. Sieve curves are considered reliable because they allow direct comparison between different grain size distributions

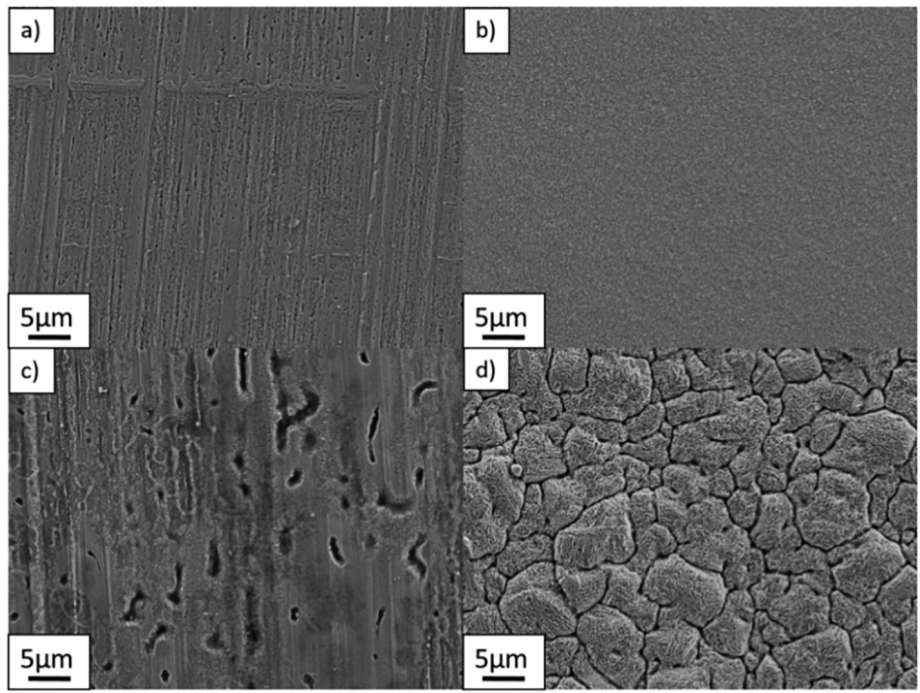

**Fig. 2.** SEM surface images of the foils investigated: a) Foil A shiny, b) Foil A matte, c) Foil B shiny, d) Foil B matte.

C.O.W. Trost, S. Wurster, C. Freitag et al.                                                                                                          Materials and Design 196 (2020) 109149 including their respective median- and average area grain sizes. The average area grain size $\overline{\nu}$ is determined as described in [44] using the following equation:

$$\overline{\nu} = \frac{\sum_{i=1}^{N} A_i \nu_i}{\sum_{i=1}^{N} A_i} \quad (1)$$

where $A_i$ is the area and $\nu_i$ the grain size of grain $i$. The sieve analysis shows that in Foil B over 40% of the grains are smaller than 1 μm whereas in Foil A only about 30% are in this regime. Foil B shows a minor part of grains larger than 4 μm up to 7 μm. Foil A shows a slight decrease in the slope of the curve at about 2.5 μm, meaning that with increasing grain size the proportions of grains in this regime becomes smaller.

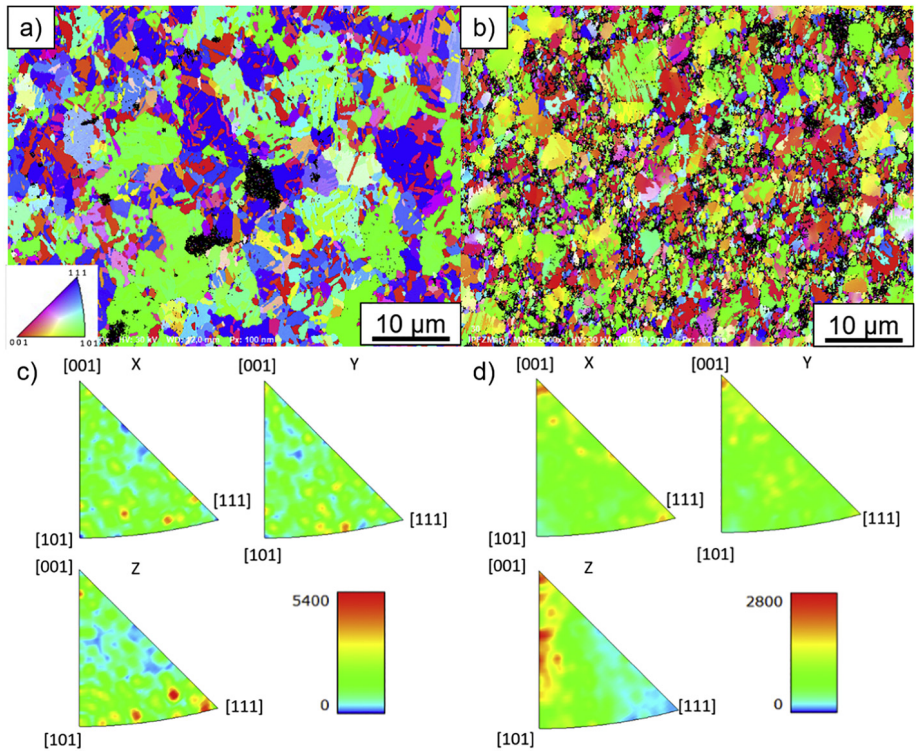

**Fig. 3.** Inverse pole figure maps of the EBSD scans and texture analysis. Y-Axis corresponds to the straining direction, X-Axis to the sample width and Z-Axis is the out-of-plane direction as described in Fig. 1. a) orientation plot of Foil A in direction Z and b) EBSD image of Foil B in direction Z. c) Inverse pole figure for Foil A and d) inverse pole figure for Foil B. Colour triangle in a) shows the colour scale for the different grain orientations in the orientation maps and valid for all EBSD orientation plots.

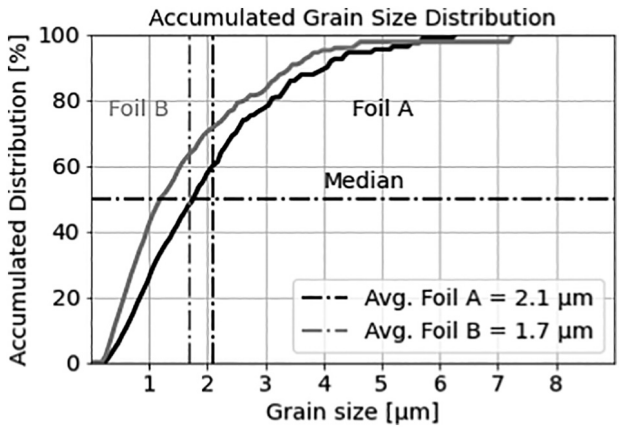

**Fig. 4.** Sieve analysis of the EBSD scans depicted in Fig. 3.

The cross sectional EBSD scans are shown in Fig. 5, with the shiny side at the top of the image and the matte side at the bottom of the figure. When analysing for Sigma 3 <111> grain boundaries one can see that Foil A shows more pronounced twinning with 1.41 μm/μm$^2$, Foil B has 0.92 μm/μm$^2$. Foil A has relatively large grains (Fig. 5a) compared to Foil B (Fig. 5b), which shows finer grains, especially in the middle of the cross section. Fig. 5c and d indicate that no strong texture was observed for either foil type. The grain size analysis in Fig. 6 shows that in Foil B over 45% of the grains are smaller than 1 μm whereas in Foil A only about 20% are in this regime. Foil A shows a minor part of grains larger than 5 μm up to 8 μm. Foil A shows a slight decrease in the slope of the curve at about 4 μm, indicating that with increasing grain size the proportions of grains in this regime become smaller.

To enable interpretation of the macro-texture using only $\theta - 2\theta$ scans (Fig. 7) the texture coefficient $M_{hkl}$ is used,

$$M_{hkl} = \frac{\frac{I_{hkl}}{I^0_{hkl}}}{\frac{1}{n}\sum \frac{I_{hkl}}{I^0_{hkl}}} \quad (2)$$

where $I_{hkl}$ is the measured peak intensity, $I^0_{hkl}$ is the intensity of the powder pattern in the ICSD – pattern [45], and $n$ is the number of peaks used for evaluation. An $M$ factor of 1 indicates randomly oriented crystallites and a factor greater 1 indicates a high amount of the specific plane. Five peaks {111}, {200}, {220}, {311} and {222} were used for the evaluation and the peak heights were obtained from the pseudo-Voigt fitting. The mean over two measurements was used. The values of $M$ corresponding to the different foils are summarized in Table 1. This method was first described [46] and has been used by others to describe thin film materials [47,48], coatings [49] as well as ED copper [44,50–52]. Differences in terms of orientation can also be seen by XRD scans presented in Fig. 7. In general, one can see that there are differences between the two sides of a foil in terms of preferred orientations with a 111 texture observed for Foil A and a 200 preferred orientation for Foil B.

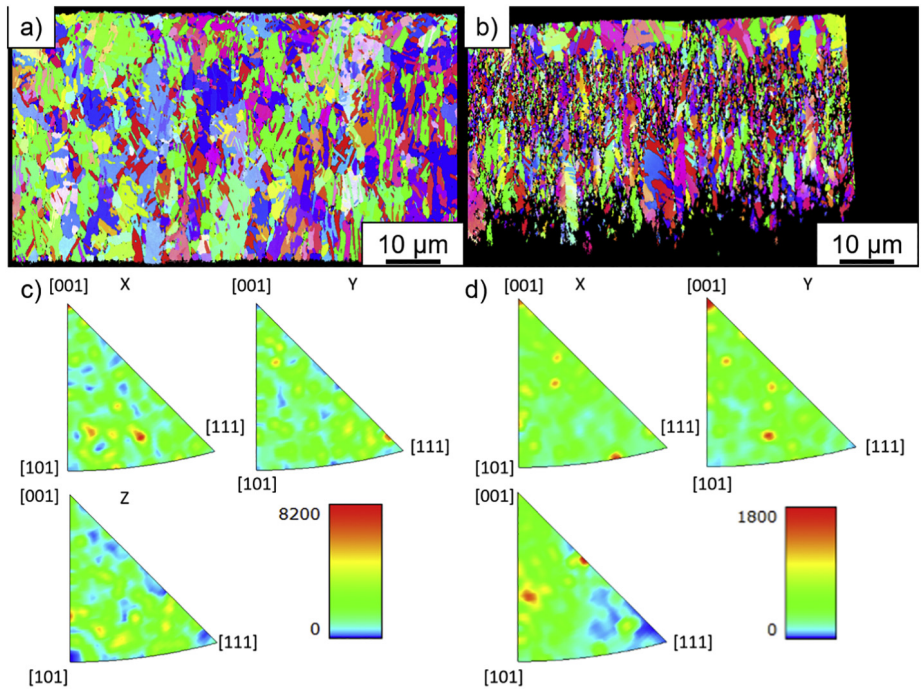

**Fig. 5.** Inverse pole figure maps of the cross sectional EBSD scans and texture analysis. Y-Axis corresponds to the straining direction, X-Axis to the sample width and Z-Axis is the out-of-plane direction as described in Fig. 1. a) Orientation plot of Foil A cross section in direction Y and b) orientation plot of Foil B cross section in direction Y. c) Inverse pole figure for Foil A and d) inverse pole figure for Foil B. The colour triangle from Fig. 3a is valid for a) and b).

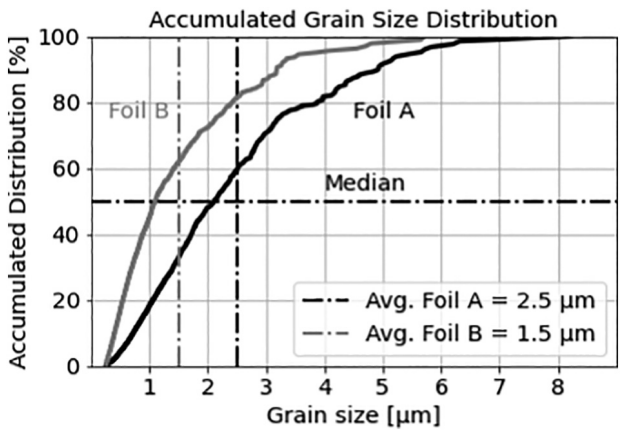

**Fig. 6.** Sieve Curve of the cross sectional EBSD scans depicted in Fig. 5.

### 3.2. Tensile Testing

Fig. 8a depicts a typical stress–strain curve for Foil A with 60 loading–unloading cycles between 20 and 87.5 N being performed in the elastic part of the curve (Fig. 8b). All curves were evaluated between 80 and 140 MPa. It can be observed that the first loading–unloading cycle response is different to the other cycles. In Fig. 8a, the transition between the linear and curved parts (elastic to plastic transition) is not always straight forward to extract. In many cases, the linear segment (rectilinear part) shows fluctuations and artefacts making it hard to extract the actual materials response. The evolution of the evaluated elastic modulus from loading and unloading portions of the 60 cycles shown in Fig. 9a and b demonstrate that after the first loading, the elastic modulus values for loading and unloading reaches a constant value. Compared to the evaluated modulus from the first loading, the fluctuations depicted in the curves are minor, thus averaging over the values is reasonable. The first loading modulus was found to be lower than the values in the plateau (constant) region as shown in Fig. 9 and Table 2. Table 2 contains the evaluation of the experiment performed on Foil A depicted in Figs. 8 and 9 and demonstrates that the elastic modulus of the first loading is significantly lower than the values calculated in the cyclic regime. As expected, the first unloading was found to be in good correlation with the mean unloading and the overall mean value over all loading and unloading cycles. Table 3 shows the averages over all experiments where six samples were evaluated per foil type. Foil A has a higher elastic modulus and overall lower deviation of the values compared to Foil B. Both foils show the same trend as described for the single experiment in Table 2. Additionally, to determine the quality of the classical evaluation of the first loading, the modulus of the first loading was evaluated for different stress regimes using a linear regression (Table 4). The elastic modulus values evaluated by the first loading differ significantly depending on the evaluated regime. The elastic modulus values of the different stress regimes show overall high standard deviations compared to the elastic modulus values evaluated by the cyclic method and the first unloading.

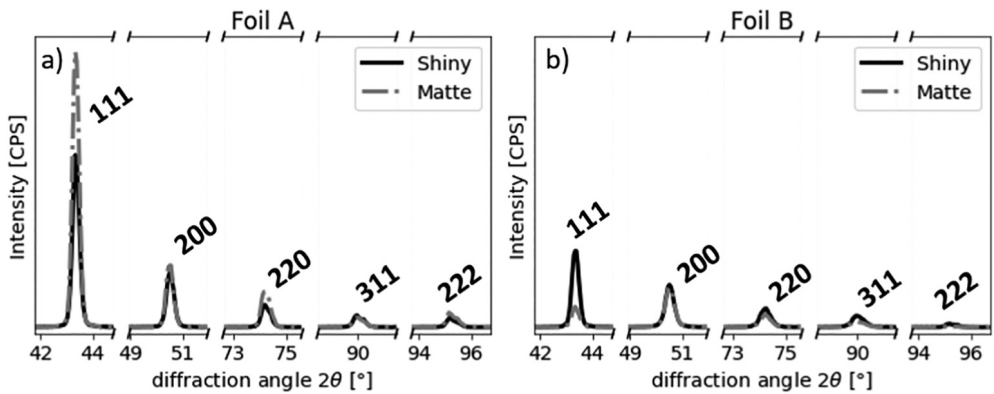

**Fig. 7.** θ − 2θ plots of the examined copper peaks for a) Foil A and b) Foil B. All peaks are shown in the same 2θ range, with the corresponding planes indicated. Both plots have the same scale in the Intensity direction.

**Table 1**
Texture coefficients $M_{hkl}$ for both foils and both sides of the corresponding foils.

|  | $M_{111}$ | $M_{200}$ | $M_{220}$ | $M_{311}$ | $M_{222}$ |
| --- | --- | --- | --- | --- | --- |
| Foil A-shiny | 1.73 | 1.19 | 0.83 | 0.34 | 0.9 |
| Foil A-matte | 1.94 | 0.95 | 0.94 | 0.21 | 0.96 |
| Foil B-shiny | 1.28 | 1.47 | 1.12 | 0.58 | 0.54 |
| Foil B-matte | 0.57 | 2.38 | 1.3 | 0.48 | 0.27 |

## 4. Discussion

SEM images of the foil surfaces can be interpreted according to [50]. The very fine surface of the shiny side of Foil A could be a result of a direct current (DC) deposition [50]. The shiny side of Foil B can be interpreted as cauliflower-like which was connected to a triangular waveform pulsed deposition of copper films [50]. The sulphur impurities in Foil B, measured by EDS can also be an indicator for a pulsed process and/or an aged bath [41,42]. It should be noted that the surface morphology should not be mistaken for the grain structure, therefore EBSD scans of the electro-polished foils were necessary [53]. When comparing the EBSD to the surface morphology of the foils one can see that although Foil A showed a finer structure at the bath side it has larger grains than Foil B.

Comparing the measured elastic moduli of both foil types one can see that they differ (see Table 3). This can be explained by the different manufacturing parameters leading to different microstructures, textures and impurity content, which ultimately lead to different mechanical responses. Differences in terms of texture can clearly be seen in Table 1 and the texture coefficient might be used for characterisation during production to ensure quality of the copper foil as reported by Bunsch et al. [52]. It is also known that the elastic modulus for Cu is highly dependent on the texture ranging between 66.7 GPa for (100) oriented grains to 191.1 GPa for (111) oriented grains [54]. Foil A ($E_{mean,overall}$ = 121 GPa) has a more pronounced (111) peak, thus a higher measured elastic modulus than Foil B (98.1 GPa) which has a more pronounced (200) peak. A high (111) peak can be attributed to the surface energy optimization [55]. This is a direct result of copper being face-centred-cubic (fcc) and grains oriented in (111) having the lowest surface energy [55]. The (111) orientation is known to be ideal for electronic applications due to its high electrical conductivity [56]. An increase in (200) orientation would result from the anisotropy of elastic constants of copper [57]. The (200) orientation is known to have the lowest strain energy, making it preferable to grow during self-annealing [55]. This could indicate that Foil B shows more pronounced self-annealing behaviour compared to Foil A. Self-annealing is known to happen in electrodeposited foils and films even at room temperature [1,55,57,58]. The self-annealing behaviour was found to depend on organic additives (brighteners) added to a sulphur based bath [55,58]. EDS showed that Foil B contained sulphur impurities, which are known to influence the self-annealing behaviour [41,42], and could lead to the microstructure of Foil B having large grains embedded in a smaller grain matrix (Fig. 5b). The importance of performing cross-sectional microstructural analysis has to be emphasized as these small grains cannot be seen in surface scans but strongly influence the mechanical behaviour of the foils. The different grain sizes directly lead to different amount of grains over the thickness (N factor) which may also play a crucial role in the mechanical behaviour of thin foils [33].

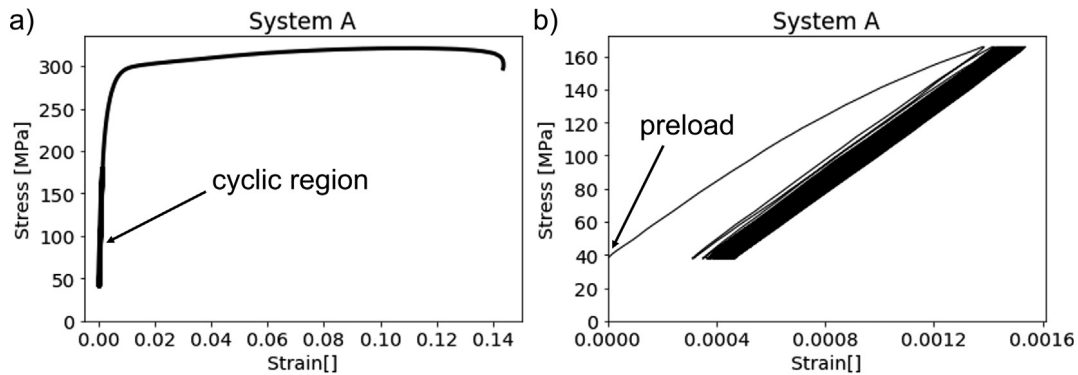

**Fig. 8.** a) Typical engineering stress–strain curve for Foil A with cyclic testing at the beginning of the test. b) Detail region of cyclic loading.

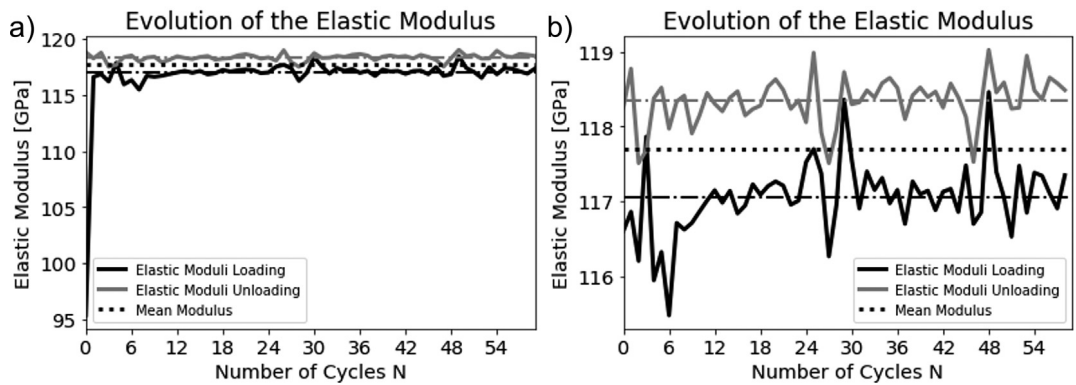

**Fig. 9.** a) Evolution of elastic modulus of Foil A (Fig. 8) and b) detail of a). The dashed lines indicate the corresponding averages in a) and b).

**Table 2**
Elastic modulus of experiments shown in Figs. 8 and 9 evaluated in a stress regime between 80 and 140 MPa.

| Elastic modulus [GPa] | First loading | First unloading | Mean loading[a] | Mean unloading | Mean overall[a] |
| --- | --- | --- | --- | --- | --- |
| Foil A – Experiment 1 | 95.3 | 118.8 | 117.1 ± 0.5 | 118.3 ± 0.3 | 117.7 ± 0.8 |

[a] Indicates that the first loading was excluded in the mean analysis.

It was shown in [3] that evaluation of the elastic modulus using a standard tensile test procedure (IPC TM-650 2.4.18.3 [31]) yields significantly lower elastic modulus values compared to using the resonance method (~factor 2–5 difference) [3]. The tests were performed on ED and rolled foils with thicknesses of 18 μm [3]. The tensile procedure was found not to be able to measure the anisotropy of the modulus and the values were found to be much lower, compared to the resonance method, which lead to reasonable values when comparing to the texture of the foils and literature data of Cu [3]. The authors concluded that the resonance method should be used instead of the tensile test procedure. Other groups [16,18] attributed rather low in-plane elastic moduli values of 20 GPa [16] and 29 GPa [18] measured via monotonic tensile tests to the microstructure. SEM analysis lead to the conclusion that the foils seemed to have columnar grains with a diameter of a few hundred nm and a high aspect ratio of about 100, when prepared via ion beam etching [18]. Similar structures were also observed after preparing cross-sections of the presented foil with ion slicing methods as shown in Fig. 10. These structures could be misinterpreted as elongated grains with a high aspect ratio, therefore in this paper EBSD analysis was performed to enable further interpretation (Fig. 5, illustrating that the microstructure was not made up of nanosized columnar grains. It has also been shown by molecular dynamics simulations on nanocrystalline copper that nanostructured grains can shift the elastic modulus to smaller values and cylindrical shaped grains enhance the effect [59], but the required grain sizes are in a regime of several nm to a few 10 nm [59]. Still other groups have calculated a grain size dependent elastic modulus of nanocrystalline copper and the respective influence of grain boundary sliding [60]. It was shown that grain boundary sliding influences the elastic modulus and increases the effect of the grain size [60], but again the necessary grain sizes were significantly smaller than the described needle-like structures in [18]. Read, et al. [14] showed that 2.6 μm thick electrodeposited copper foils resembled an agglomeration of round balls or spheres of approximately 30–50 nm, although EBSD found grain sizes of >1 μm and XRD crystallite sizes of few 100 nm [14]. They were able to show via molecular dynamics calculations that the morphology of the foils in [14] resulted in Young's modulus values tens of percent lower than bulk values [61].

As a result of the presented calculations and the fact that the preparation by ion slicing methods can lead to artefacts, it has to be concluded that the severe differences in the measured elastic modulus should not be purely attributed to microstructural differences or grain boundary sliding, but also to the described problems with standard tensile test evaluation procedures.

The reproducibility of the evaluation of loading–unloading cycles was first demonstrated for thin freestanding films [6,7]. It was shown that loading–unloading cycles in free-standing thin films of gold and copper with thicknesses ranging from 0.05 μm to 2 μm resulted in a reproducible slope with either decreasing or increasing stress. When evaluating the elastic modulus from the engineering stress–strain data of the foils studied here, it is very important to distinguish between the straight and curved parts of the loading–unloading curve. Hwangbo and Song [9] used a regression analysis to calculate the elastic modulus. In the current study presented here another approach to find the optimal range for evaluation was used and accomplished by plotting the gradient of stress over the gradient of strain for each data point (loading or unloading segment) of the initially rectilinear (straight) portion of the curve. The calculated value resembles the local tangent modulus of the stress–strain curve. In a perfect test without any measurement fluctuations of an ideally elastic material the local tangent modulus will show a constant value over a significant stress range. When the same approach is applied to the first loading segment of the stress–strain curves measured for Foils A and B, a constant value was not observed (Fig. 11a), thus, making it impossible to distinguish the curved or straight parts of the stress–strain curve from possible artefacts in the stress–strain curve. With repetitive loading, the evaluation of the data produces gradient plots for loading and unloading segments which show a plateau that clearly corresponds to the linear parts of the loading or unloading cycle (Fig. 11b–f). Additionally, the fluctuations in the regression analysis curves were found to decrease with increasing cycle number (Fig. 11c and f). These plateaus were used to estimate the optimal boundary values for the evaluation with a regression analysis. A value of 80–140 MPa was found to be ideal, by plotting all respective curves and analysing them for the plateau. This value is in good correlation with literature [9] which found the cyclic stress range of perfect elastic behaviour of 12 μm ED Cu-foils to be between 85 and 115 MPa. By analysing plateaus over a number of cycles one can see that the value of the modulus in Fig. 9a shows no significant changes over cycle number after the first loading segment, which is in agreement with the small standard deviation of elastic moduli shown in Table 2. When comparing the first unloading with the overall mean in Table 3 and the standard deviation of one experiment it is demonstrated that in practice it would be sufficient to simply take the first unloading segment. Thus, the general experimental procedure would include first loading the sample, then unloading while still in the elastic regime, and loading again to failure. However, for achieving higher statistical significance of the result, higher cycle numbers are advised. The unloading segment is necessary because during the first loading, the foil "curtains" (makes ripples or folds similar to a window curtain), which causes the artefacts in the initial loading curve. By unloading and loading again, the curtaining is removed.

**Table 3**
Averages of the elastic modulus over all experiments evaluated in a stress regime between 80 and 140 MPa.

| Elastic Modulus [GPa] | First loading | First unloading | Mean loading[*] | Mean unloading | Mean overall[*] |
| --- | --- | --- | --- | --- | --- |
| Foil A | 100.3 ± 5.8 | 122.4 ± 3.1 | 120.5 ± 3.3 | 121.5 ± 2.9 | 121 ± 3.1 |
| Foil B | 78.3 ± 10.6 | 100.7 ± 8.3 | 97.4 ± 6.4 | 98.8 ± 7.5 | 98.1 ± 6.5 |

[*] Indicates that the first loading was excluded in the mean analysis.

**Table 4**
First loading averages evaluated in different stress regimes of the experiments evaluated in Table 3.

| First loading | Elastic Modulus [GPa] evaluated in stress regime of | | |
|---|---|---|---|
| | 40–60 MPa | 40–80 MPa | 40–100 MPa |
| Foil A | 111.5 ± 26.1 | 115.3 ± 12.4 | 115.7 ± 7.4 |
| Foil B | 104.2 ± 26.5 | 96.9 ± 18.6 | 90.8 ± 13.7 |

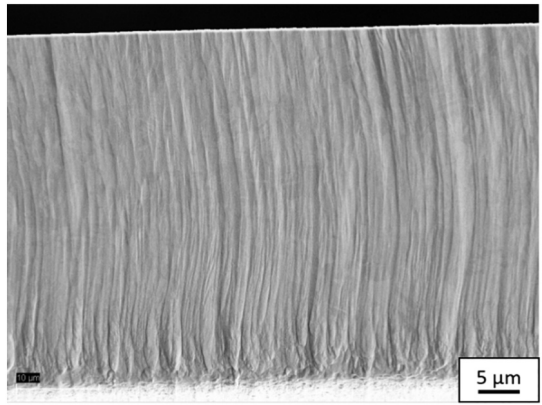

**Fig. 10.** Ion polished Foil A showing artefacts as a result of preparation via ion polishing, same sample as depicted in Fig. 5a).

When comparing the classical first loading evaluation of Table 4 with the values obtained by the cyclic and the first unloading method (Table 3), one can see that depending on the stress regime (Table 4) the evaluated moduli vary significantly. Also the standard deviations of the measured elastic moduli are significantly higher than in the cyclic experiments showing that the described influences in the first loading regime make it impossible to extract satisfactory materials data. The high ambiguity of the evaluation of the first loading regime and the high deviation from the moduli derived from cyclic loading, show that this kind of evaluation can lead to dramatically different, physically unrealistic elastic modulus values even for the same foil. Therefore, the proposed cyclic and the unloading method are considered as an improvement to all standard tensile procedures for testing thin foils.

The results presented illustrate that evaluation of the elastic properties of copper foils using only the first straining part of the curves can lead to severe differences compared to other methods and theoretical texture based calculations. The high anisotropy of the elastic modulus of copper ranging from 66.7 GPa for (100) oriented copper [54] to 191.1 GPa for (111) oriented copper [54] combined with monotonic tensile tests can easily lead to inaccurate values. It should be kept in mind that these values are only valid for single crystals, so the actual elastic modulus both for random and textured microstructures should lie between those values, if not influenced by other factors such as nanostructured grains, morphology and grain boundary sliding [59–61]. Values significantly lower should be handled with care. The method proposed here allows controlled evaluation of elastic properties of foils, minimizing experimental influences such as curtaining.

## 5. Conclusion

It was shown on two 40 μm thick electrodeposited copper foils that evaluation of the elastic modulus is not trivial for metallic foils. The importance of a controlled procedure was emphasized using experiments and comparisons to literature. Measurements have to be performed in a way that minimizes influences of other effects not related to the purely elastic behaviour of the material. Therefore, a cyclic tensile straining approach in the elastic regime was introduced and evaluated using the local tangent moduli as an indicator of the most linear segment of the stress–strain curve, enabling the measurement of the elastic modulus with high accuracy. As a result of the different deposition methods of the foils, different microstructures and textures were achieved leading to different elastic modulus values. The evaluated elastic modulus for Foil A was found to be 120 GPa and for Foil B slightly lower at 98 GPa. The lower value is most likely due to the differences in texture. Both foils have reasonable elastic modulus values for polycrystalline Cu. It was also demonstrated that the use of at least one unloading segment to measure the elastic modulus is a significant improvement over no unloading in the elastic regime in order to remove the curtaining effect. With this technique no editing or removing parts of the data is needed to guarantee a physically valid evaluation. The new approach should be considered to accurately measure the elastic modulus of foil and ribbon materials.

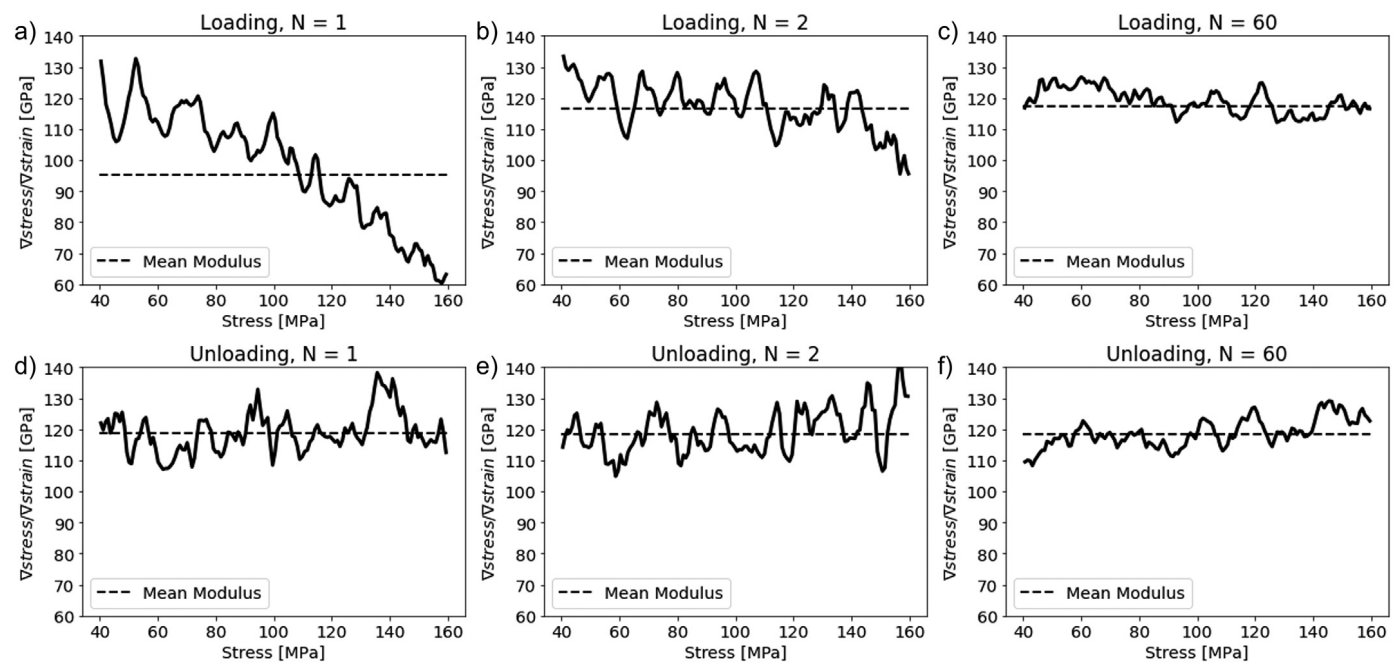

**Fig. 11.** a)–f) Evolution of the $\frac{\nabla stress}{\nabla strain}$ of Foil A (Fig. 8) for the corresponding cycle, N.

**Declaration of Competing Interest**

The authors declare that they have no known competing financial interests or personal relationships that could have appeared to influence the work reported in this paper.

**Acknowledgement**

The authors gratefully acknowledge the financial support under the scope of the COMET program within the K2 Center "Integrated Computational Material, Process and Product Engineering (IC-MPPE)" (Project No 859480). This program is supported by the Austrian Federal Ministries for Climate Action, Environment, Energy, Mobility, Innovation and Technology (BMK) and for Digital and Economic Affairs (BMDW), represented by the Austrian Research Funding Association (FFG), and the federal states of Styria, Upper Austria and Tyrol. This project received funding from the European Research Council (ERC) under the European Union's Horizon 2020 research and innovation programme (Grant No. 757333).

**References**

[1] A.A. Volinsky, et al., Microstructure and mechanical properties of electroplated cu thin films, MRS Proc. 649 (2000) 174, https://doi.org/10.1557/PROC-649-Q5.3.

[2] H.D. Merchant, G. Khatibi, B. Weiss, Elastic and elastoplastic response of thin copper foil, J. Mater. Sci. 39 (13) (2004) 4157–4170, https://doi.org/10.1023/B:JMSC.0000033395.87373.ea.

[3] K. Kammuri, A. Miki, H. Takeuchi, Reliable Young's Modulus value of high flexible, treated rolled copper foils measured by resonance method, J. Microelectron. Electron. Packag. 14 (2) (2017) 70–76, https://doi.org/10.4071/imaps.454688.

[4] J. Zhu, J. Feng, Z. Guo, Mechanical properties of commercial copper current-collector foils, RSC Adv. 4 (101) (2014) 57671–57678, https://doi.org/10.1039/C4RA07675C.

[5] J.-M. Song, D.-S. Wang, C.-H. Yeh, W.-C. Lu, Y.-S. Tsou, S.-C. Lin, Texture and temperature dependence on the mechanical characteristics of copper electrodeposits, Mater. Sci. Eng. A 559 (2013) 655–664, https://doi.org/10.1016/j.msea.2012.09.006.

[6] C.A. Neugebauer, Tensile properties of thin, evaporated gold films, J. Appl. Phys. 31 (6) (1960) 1096–1101, https://doi.org/10.1063/1.1735751.

[7] C.A.O. Henning, F.W. Boswell, J.M. Corbett, Mechanical properties of vacuum-deposited metal films—I. copper films, Acta Metall. 23 (2) (1975) 177–185, https://doi.org/10.1016/0001-6160(75)90181-9.

[8] A. Lawley, S. Schuster, Tensile behavior of copper foils prepared from rolled material, Trans. Metall. Soc. AIME 230 (1) (1964) 27–33.

[9] Y. Hwangbo, J.-H. Song, Fatigue life and plastic deformation behavior of electrodeposited copper thin films, Mater. Sci. Eng. A 527 (9) (2010) 2222–2232, https://doi.org/10.1016/j.msea.2010.01.016.

[10] D.T. Read, Y.-W. Cheng, R. Geiss, Mechanical behavior of electrodeposited copper film at elevated temperatures, Proceedings of the ASME Materials Division - 2004: Presented at 2004 ASME International Mechanical Engineering Congress and Exposition, November 13–19, 2004, Anaheim, California, USA, Anaheim, California, USA 2004, pp. 113–116.

[11] D. Read, Tension-tension fatigue of copper thin films, Int. J. Fatigue 20 (3) (1998) 203–209, https://doi.org/10.1016/S0142-1123(97)00080-7.

[12] H. Huang, F. Spaepen, Tensile testing of free-standing cu, Ag and Al thin films and Ag/cu multilayers, Acta Mater. 48 (12) (2000) 3261–3269, https://doi.org/10.1016/S1359-6454(00)00128-2.

[13] D.T. Read, Y.-W. Cheng, R.R. Keller, J.D. McColskey, Tensile properties of free-standing aluminum thin films, Scr. Mater. 45 (5) (2001) 583–589, https://doi.org/10.1016/S1359-6462(01)01067-3.

[14] D.T. Read, Y.W. Cheng, R. Geiss, Morphology, microstructure, and mechanical properties of a copper electrodeposit, Microelectron. Eng. 75 (1) (2004) 63–70, https://doi.org/10.1016/j.mee.2003.09.012.

[15] S. Zhang, M. Sakane, T. Nagasawa, K. Kobayashi, Mechanical properties of copper thin films used in electronic devices, Procedia Eng. 10 (2011) 1497–1502, https://doi.org/10.1016/j.proeng.2011.04.250.

[16] N. Murata, K. Tamakawa, H. Miura, Fatigue strength of electroplated copper thin films under uni-axial stress, Proc. Mater. Mech. Conf. 2007 (0) (2007) 173–174, https://doi.org/10.1299/jsmemm.2007.173.

[17] J.-H. Park, J.-H. An, Y.-J. Kim, Y.-H. Huh, H.-J. Lee, Tensile and high cycle fatigue test of copper thin film, Mat.-wiss. u. Werkstofftech. 39 (2) (2008) 187–192, https://doi.org/10.1002/mawe.200700262.

[18] K. Tamakawa, K. Sakutani, H. Miura, Effect of micro texture of electroplated copper thin films on their mechanical properties, J. Soc. Mat. Sci., Japan 56 (10) (2007) 907–912, https://doi.org/10.2472/jsms.56.907.

[19] B. Weiss, et al., Characterization of mechanical and thermal properties of thin cu foils and wires, Sensors Actuators A Phys. 99 (1) (2002) 172–182, https://doi.org/10.1016/S0924-4247(01)00877-9.

[20] R.L. Grunes, C. D'Antonio, F.K. Kies, Mechanical properties of thin nickel films, J. Appl. Phys. 36 (9) (1965) 2735–2739, https://doi.org/10.1063/1.1714570.

[21] J.A. Ruud, D. Josell, F. Spaepen, A.L. Greer, A new method for tensile testing of thin films, J. Mater. Res. 8 (1) (1993) 112–117, https://doi.org/10.1557/JMR.1993.0112.

[22] A. Wimmer, et al., Damage evolution during cyclic tension–tension loading of micron-sized Cu lines, Acta Mater. 67 (2014) 297–307, https://doi.org/10.1016/j.actamat.2013.12.006.

[23] H.-G. Min, D.-J. Kang, J.-H. Park, Comparison of Tensile and Fatigue Properties of Copper Thin Film Depending on Process Method, Appl. Sci. 10 (1) (2020) 388, https://doi.org/10.3390/app10010388.

[24] S.H. Hong, K.S. Kim, Y.-M. Kim, J.-H. Hahn, C.-S. Lee, J.-H. Park, Characterization of elastic moduli of Cu thin films using nanoindentation technique, Compos. Sci. Technol. 65 (9) (2005) 1401–1408, https://doi.org/10.1016/j.compscitech.2004.12.010.

[25] J. Káňa, B. Mašek, K. Rubešová, Measuring material properties of metal foils using bulge test method, Procedia Eng. 100 (2015) 861–867, https://doi.org/10.1016/j.proeng.2015.01.442.

[26] M.K. Small, W.D. Nix, Analysis of the accuracy of the bulge test in determining the mechanical properties of thin films, J. Mater. Res. 7 (6) (1992) 1553–1563, https://doi.org/10.1557/JMR.1992.1553.

[27] Y. Xiang, X. Chen, J.J. Vlassak, The mechanical properties of electroplated cu thin films measured by means of the bulge test technique, MRS Proc. 695 (2001) 171, https://doi.org/10.1557/PROC-695-L4.9.1.

[28] P.R. Cantwell, et al., Estimating the in-plane Young's modulus of polycrystalline films in MEMS, J. Microelectromech. Syst. 21 (4) (2012) 840–849, https://doi.org/10.1109/JMEMS.2012.2191939.

[29] M. Klein, A. Hadrboletz, B. Weiss, G. Khatibi, The 'size effect' on the stress–strain, fatigue and fracture properties of thin metallic foils, Mater. Sci. Eng. A 319-321 (2001) 924–928, https://doi.org/10.1016/S0921-5093(01)01043-7.

[30] A. Catlin, W.P. Walker, Mechanical properties of thin single-crystal gold films, J. Appl. Phys. 31 (12) (1960) 2135–2139, https://doi.org/10.1063/1.1735513.

[31] IPC-TM, IPC-TM-650Test Methods Manual, 2020 https://www.ipc.org/TM/2.4.18.3.pdf 2.4.18.3. (7/95).

[32] ASTM E345-16, Standard Test Methods of Tension Testing of Metallic Foil, ASTM International, West Conshohocken, PA, 2016.

[33] M.W. Fu, W.L. Chan, Geometry and grain size effects on the fracture behavior of sheet metal in micro-scale plastic deformation, Mater. Des. 32 (10) (2011) 4738–4746, https://doi.org/10.1016/j.matdes.2011.06.039.

[34] D.F. Bahr, S.L. Jennerjohn, D.J. Morris, Dislocation nucleation and multiplication in small volumes: the onset of plasticity during indentation testing, JOM 61 (2) (2009) 56–60, https://doi.org/10.1007/s11837-009-0029-3.

[35] Hossein Alimadadi, Karen Pantleon, Challenges of sample preparation for cross sectional EBSD analysis of electrodeposited nickel films, 2009 183–189.

[36] S.-H. Kim, J.-H. Kang, S.Z. Han, Electron backscatter diffraction characterization of microstructure evolution of electroplated copper film, Mater. Trans. 51 (4) (2010) 659–663, https://doi.org/10.2320/matertrans.MG200910.

[37] J. Mikeš, S. Pekárek, O. Babčenko, O. Hanuš, J. Kákona, P. Štenclová, 3D printing materials for generators of active particles based on electrical discharges, Plasma Process. Polym. 17 (1) (2020) 1900150, https://doi.org/10.1002/ppap.201900150.

[38] M.D. Womble, J. Herbsommer, Y.-J. Lee, J.W.P. Hsu, Effects of TiO2 nanoparticle size and concentration on dielectric properties of polypropylene nanocomposites, J. Mater. Sci. 53 (12) (2018) 9149–9159, https://doi.org/10.1007/s10853-018-2223-6.

[39] S. Soltani, P. S. Taylor, and J. C. Batchelor, "Mechanically influenced antennas for strain sensing applications using multiphysics modelling," in EuCAP 2020: 14th European Conference on Antennas and Propagation: 15–20 March 2020, Copenhagen, Denmark, 2020, pp. 1–4.

[40] R.M. Joseph, et al., Synthesis and characterization of polybenzimidazole membranes for gas separation with improved gas permeability: a grafting and blending approach, J. Membr. Sci. 564 (2018) 587–597, https://doi.org/10.1016/j.memsci.2018.07.064.

[41] L.T. Koh, G.Z. You, C.Y. Li, P.D. Foo, Investigation of the effects of byproduct components in cu plating for advanced interconnect metallization, Microelectron. J. 33 (3) (2002) 229–234, https://doi.org/10.1016/S0026-2692(01)00122-7.

[42] M. Stangl, et al., Incorporation of sulfur, chlorine, and carbon into electroplated cu thin films, Microelectron. Eng. 84 (1) (2007) 54–59, https://doi.org/10.1016/j.mee.2006.08.004.

[43] K. Mannesson, M. Elfwing, A. Kusoffsky, S. Norgren, J. Ågren, Analysis of WC grain growth during sintering using electron backscatter diffraction and image analysis, Int. J. Refract. Met. Hard Mater. 26 (5) (2008) 449–455, https://doi.org/10.1016/j.ijrmhm.2007.10.004.

[44] G.-T. Lui, D. Chen, J.-C. Kuo, EBSD characterization of twinned copper using pulsed electrodeposition, J. Phys. D Appl. Phys. 42 (21) (2009) 215410, https://doi.org/10.1088/0022-3727/42/21/215410.

[45] I. Levin, NIST Inorganic Crystal Structure Database (ICSD), 2020.

[46] G.B. Harris, X. Quantitative measurement of preferred orientation in rolled uranium bars, The London, Edinburgh, and Dublin Philosophical Magazine and Journal of Science 43 (336) (1952) 113–123, https://doi.org/10.1080/14786440108520972.

[47] G. Singh, S.B. Shrivastava, D. Jain, S. Pandya, T. Shripathi, V. Ganesan, Effect of indium doping on zinc oxide films prepared by chemical spray pyrolysis technique, Bull. Mater. Sci. 33 (5) (2010) 581–587, https://doi.org/10.1007/s12034-010-0089-6.

[48] T. Prasada Rao, M.C. Santhoshkumar, Highly oriented (100) ZnO thin films by spray pyrolysis, Appl. Surf. Sci. 255 (16) (2009) 7212–7215, https://doi.org/10.1016/j.apsusc.2009.03.065.

[49] L. Feng, Y.-Y. Ren, Y.-h. Zhang, S. Wang, L. Li, Direct Correlations among the Grain Size, Texture, and Indentation Behavior of Nanocrystalline Nickel Coatings, Metals 9 (2) (2019) 188, https://doi.org/10.3390/met9020188.

[50] A. Ibañez, E. Fatás, Mechanical and structural properties of electrodeposited copper and their relation with the electrodeposition parameters, Surf. Coat. Technol. 191 (1) (2005) 7–16, https://doi.org/10.1016/j.surfcoat.2004.05.001.

[51] Q. Liao, L.-Q. Zhu, H.-C. Liu, W.-P. Li, Mechanical properties of electroformed copper layers with gradient microstructure, Int. J. Miner. Metall. Mater. 17 (1) (2010) 69–74, https://doi.org/10.1007/s12613-010-0112-3.

[52] A. Bunsch, S.J. Skrzypek, J. Kowalska, W. Ratuszek, W. Rakowski, Texture and mechanical properties of electrodeposited copper thin films, SSP 163 (2010) 141–144, https://doi.org/10.4028/www.scientific.net/SSP.163.141.

[53] H.D. Merchant, W.C. Liu, L.A. Giannuzzi, J.G. Morris, Grain structure of thin electrodeposited and rolled copper foils, Mater. Charact. 53 (5) (2004) 335–360, https://doi.org/10.1016/j.matchar.2004.07.013.

[54] A. Basavalingappa, J.R. Lloyd, Effect of microstructure and anisotropy of copper on reliability in Nanoscale interconnects, IEEE Trans. Device Mater. Relib. 17 (1) (2017) 69–79, https://doi.org/10.1109/TDMR.2017.2655459.

[55] M. Hasegawa, Y. Nonaka, Y. Negishi, Y. Okinaka, T. Osaka, Enhancement of the Ductility of Electrodeposited Copper Films by Room-Temperature Recrystallization, J. Electrochem. Soc. 153 (2) (2006) C117, https://doi.org/10.1149/1.2149299.

[56] B.V. Sarada, C.L.P. Pavithra, M. Ramakrishna, T.N. Rao, G. Sundararajan, Highly (111) Textured Copper Foils with High Hardness and High Electrical Conductivity by Pulse Reverse Electrodeposition, Electrochem. Solid-State Lett. 13 (6) (2010) D40, https://doi.org/10.1149/1.3358145.

[57] H. Lee, S.S. Wong, S.D. Lopatin, Correlation of stress and texture evolution during self- and thermal annealing of electroplated Cu films, J. Appl. Phys. 93 (7) (2003) 3796–3804, https://doi.org/10.1063/1.1555274.

[58] S. Lagrange, et al., Self-annealing characterization of electroplated copper films, Microelectron. Eng. 50 (1) (2000) 449–457, https://doi.org/10.1016/S0167-9317(99)00314-7.

[59] K. Zhou, B. Liu, Y. Yao, K. Zhong, Effects of grain size and shape on mechanical properties of nanocrystalline copper investigated by molecular dynamics, Mater. Sci. Eng. A 615 (2014) 92–97, https://doi.org/10.1016/j.msea.2014.07.066.

[60] P. Sharma, S. Ganti, On the grain-size-dependent elastic modulus of nanocrystalline materials with and without grain-boundary sliding, J. Mater. Res. 18 (8) (2003) 1823–1826, https://doi.org/10.1557/JMR.2003.0253.

[61] D.T. Read, Atomistic simulation of modulus deficit in an aggregate of metal spheres, J. Appl. Phys. 97 (1) (2005) 13522, https://doi.org/10.1063/1.1819978.